\documentclass[12pt]{article}
\usepackage{amssymb}
\usepackage{epsfig}
\usepackage{graphicx}
\usepackage{amsmath}
\usepackage{verbatim}

\csname @addtoreset\endcsname{equation}{section}
\begin{document}
\begin{titlepage}
\title{\bf\Large A Note on Bounds of Scalar Operators in Perturbative SCFTs \vspace{18pt}}

\author{\normalsize Sibo~Zheng \vspace{12pt}\\
{\it\small   Department of Physics, Chongqing University, Chongqing 401331, P.R. China}\\}

\date{}
\maketitle \voffset -.3in \vskip 1.cm \centerline{\bf Abstract}
\vskip .3cm  Bounds on anomalous dimensions of scalar operators in 4d superconformal field theory are explored through perturbative viewpoint.
Following the recent work of Green and Shih,  in which a conjecture involved this issue is verified at the NLO,
we consider the NNLO corrections to the bounds, which are important in some situations and can be divided into two cases
where $\mathcal{O}(\lambda^{4})$ or $\mathcal{O}(y^2) $ effects dominate respectively.
In the former case, we find that the conjecture is maintained at NNLO,
while in the later case the statement still holds due to null correction.

\vskip 5.cm \noindent May 2012
 \thispagestyle{empty}

\end{titlepage}
\section{Introduction}
Conformal field theory (CFT) ( see \cite{Mack, DP, Osborn} for example ),
which is tied to important concepts in field theory and phenomenological application,
has been extensively explored.
For example, the small hierarchy $\mu$ problem
involved in electroweak symmetry breaking in the minimal supersymmetric standard model
can be solved when this theory is coupled to a \em{hidden} \em  superconformal field theory (SCFT) \cite{mu1, mu2}
(for other recipes, see \cite{Zheng} and reference therein for example ).
The reason for this viability is due to the different scaling behaviors between chiral $\mu$ and real $B_{\mu}$ operator,
which is expected in  SCFTs where the condition $\delta_{min}>0$ (see its definition in \eqref{E4}) is satisfied.

Given a CFT,  the dimensions of operators and coefficients in the correlator functions
(or equivalently the OPE coefficients) of these operators
exactly \em{determine}\em~ or \em{define} \em the theory.
Many efforts have been done  by using arguments of conformal symmetry, crossing symmetry and unitarity.
Among these developments,
an interesting and well-known topic in unitary CFT is the discovery of bounds on dimensions of operators.
The full list of unitary bounds, which includes fields with Lorentz spin $(j, \tilde{j})$ is presented in \cite{Mack} .
Also, a-maximization \cite{a} that follows from the arguments involved in anomalies of global symmetries
provides, in terms of unitary constraints,
an alternative method to determine the dimensions of chiral operators in SCFTs.

Very recently the bounds on anomalous dimension of primary scalar operators are addressed
\cite{1106.4037,0807.0004, 1009.2725, 0912.2726, 1009.5985, 1009.2087, 1109.5176}
by applying  conformal blocks \cite{Osborn2} and global symmetries to
exploring the four-point correlators of scalar primary operators.
A conjecture is hinted by these works.

The 4d interacting SCFT $\mathcal{P}_1$ we are going to study contains a chiral operator $\mathcal{O}$ of dimension $\Delta_{\mathcal{O}}=2-\epsilon$.
The OPE of $\mathcal{O}$ and its anti-chiral field $\mathcal{O}^{\dag}$ is assumed to be
\begin{eqnarray}{\label{E1}}
\mathcal{O}(x)^{\dag}\mathcal{O}(0)=\frac{1}{\mid x\mid^{2\Delta_{\mathcal{O}}}}+\sum_{i}\frac{c_{i}}{\mid x\mid^{2\Delta_{\mathcal{O}}-\Delta_{i}}}L_{i}+\cdots
\end{eqnarray}
where $L_{i}$ are real scalar multiplets with dimension $\Delta_{i}=2+\nu_{i}$ ( Here $\nu_{i}$ is a non-negative real number ).
$c_i$ refer to the OPE coefficients.
The terms ignored in \eqref{E1} denote descendants with higher spin.
We follow the convention in \cite{1203.5129} where all primary scaling operators are canonically normalized as in  \eqref{E1}.
We explore theories constructed through deforming $\mathcal{P}_1$ by ,
\begin{eqnarray}{\label{E2}}
\mathcal{L}=\mathcal{L}_{\mathcal{P}_{1}}+\left(\frac{1}{4\pi^{2}}\int d^{4}\theta X^{\dag}X +\int d^{2}\theta \frac{\lambda}{2\pi}X\mathcal{O}+h.c \right)
\end{eqnarray}
with $X$ being a free chiral superfield.
Our concern is to discuss the anomalous dimensions of scalar primary operators $\mathcal{S}_{i}$
which appear in the OPE of $X$ and $X^{\dag}$,
\begin{eqnarray}{\label{E3}}
X^{\dag}(x)X(0)=\frac{1}{\mid x\mid^{2\Delta_{X}}}+\sum_{i} \frac{c_{i}}{\mid x\mid^{2\Delta_{X}-\Delta_{i}}}\mathcal{S}_{i}+\cdots,
\end{eqnarray}
When  the anomalous dimension of $\Delta_{X}=1+\epsilon$ is small, $0 <\epsilon<<1$  as we assume throughout this paper,
the deformed theory \eqref{E2} will renormalization group (RG) flow into a new interacting CFT $\mathcal{P}_{2}$.
As expected, the candidate operators $\mathcal{S}_{i}$ in \eqref{E3} include $X^{\dag}X$,  $L_{i}$ and their mixing.
A variety of works \cite{1106.4037,0807.0004, 1009.2725, 0912.2726, 1009.5985, 1009.2087, 1109.5176}
tend to claim that the sign of $\delta_{min}$ defined as
\footnote{As mentioned in the previous discussion, the study of this conjecture is of interest from point of view of phenomenology.}
\begin{eqnarray}{\label{E4}}
\delta_{min} =\min{(\Delta_{i})}-2\Delta_{X}<0
\end{eqnarray}
always holds in general.

The purpose in this article is to study the higher-order corrections on this conjecture in the context of perturbative CFT,
by following the method of calculations proposed  by Green and Shih \cite{1203.5129}.
The advantage of this method is that the RG flow  between the new and old fixed points is manifest.
By using this method,  the conjecture is perturbatively verified at the next-to-leading order ( NLO).
We would like to address the question whether the the bound on $\delta_{min}$ is robust
as suggested.
If not, then under which circumstances it can be violated.
As we will claim,  despite smaller than NLO ones,
the NNLO corrections are important and even substantial in some circumstances.
In particular, the modifications to the vanishing matrix elements of anomalous dimension of $\mathcal{S}_i$ at NLO
can directly affect the sign of $\delta_{min}$,
even though they don't substantially modify the values of fixed points couplings $\lambda_{*}$ and $y_{i *}$.

In section 2,  we divide the discussions into two cases.
In the case  where $\mathcal{O}(\lambda^{4})$ dominates,
we calculate the corrections to values of couplings at the new fixed points in section 3,
and estimate the modification to the matrix of anomalous dimension and value of $\delta_{min}$,
which are found to be substantial, however, not enough to violate the conjecture.
In section 4, we consider the modification due to $\mathcal{O}(y^{2})$ effects at NNLO,
which is found to be actually \em{null}\em.
We claim that this observation \em{exactly}\em~ holds beyond NLO.
Finally , we summarize our results in section 5.

\section{NNLO Corrections}
Take the RG effects into account, the Lagrangian for $\mathcal{P}_{2}$ SCFT can be written as,
\begin{eqnarray}{\label{D1}}
\mathcal{L}=\mathcal{L}_{\mathcal{P}_{1}}+\frac{1}{4\pi^{2}}\int d^{4}\theta (1+\delta Z_{X})X^{\dag}X + \int d^{4}\theta
(y_{i}+\delta y_{i})L_{i}+\left(\int d^{2}\theta \frac{\lambda}{2\pi}\Lambda^{\epsilon}X\mathcal{O}+h.c \right) \nonumber\\
\end{eqnarray}
where we have introduced $\Lambda$ dependence so that $\lambda$ is a dimensionless coupling.
$y_i$ are the coupling constants appearing in $L_i$ operators.
$\delta Z_{X}$ and $\delta~y_{i}$ denote the effects of wave-function renormalization.
By using the holomorphic arguments,
we find the beta function for $\lambda$ is exactly given by,
\begin{eqnarray}{\label{D2}}
\beta_{\lambda}=-\epsilon\lambda +\lambda \gamma_{X}(\lambda, y_{i}),~~~~~~
\gamma_{X}=-\frac{1}{2}\frac{\partial \delta Z_{X}}{\partial \log\Lambda}
\end{eqnarray}
Expanding the wave-function renormalization functionals $\delta Z_{X}$ and $\delta y_{i}$ in power of $\lambda$ and $y_{i}$ which are both
assumed to be small as,
\begin{eqnarray}{\label{D3}}
\delta Z_{X}&=&a_{1}\lambda^{2}+a_{1i}y_{i}+a_{2i}\lambda^{2}y_{i}+ a_{2}\lambda^{4}+a_{2ij}y_{i}y_{i}+\mathcal{O}(\lambda^{6}, y^{4}, \lambda^{4}y^{2}) \nonumber\\
\delta y_{i}&=&b_{1i}\lambda^{2}+b_{1ij}y_{j}+b_{2ij}\lambda^{2}y_{j}+b_{2i}\lambda^{4}+b_{2ijk}y_{j}y_{k}+\mathcal{O}(\lambda^{6}, y^{4}, \lambda^{4}y^{2})
\end{eqnarray}
where $a_i$, $b_i$ are real coefficients,
some of which have been considered in \cite{1203.5129} up to NLO,
\begin{eqnarray}{\label{D4}}
a_{1~}&=&\frac{\pi^{2}}{\epsilon},\nonumber\\
a_{1i}&=&0 ,\\
a_{2i}&=&\frac{8\pi^{4}c_{i}}{\nu_{i}-2\epsilon}\mathcal{I}(\nu_{i},\epsilon),\nonumber
\end{eqnarray}
and
\begin{eqnarray}{\label{D5}}
b_{1i~}&=&\frac{c_{i}}{2(2\epsilon+\nu_{i})},\nonumber\\
b_{1ij}&=&0,\\
b_{2ij}&=&0,\nonumber
\end{eqnarray}

In the following we take into account the NNLO corrections.
In terms of the assumption in \eqref{D3} we can write the beta function of  $\lambda$ and $y_i$ as,
\begin{eqnarray}{\label{B1}}
\beta_{\lambda}&=&-\epsilon\lambda+\lambda\left[\pi^{2}\lambda^{2}-4\pi^{4}\sum_{i}c_{i}y_{i}\mathcal{I}(\nu_{i}, \epsilon) \lambda^{2}+2\epsilon a_{2}\lambda^{4}-\sum_{i, j} a_{2ij}(\nu_{i}+\nu_{j})y_{i}y_{j}\right]\nonumber\\
\beta_{y_{i}}&=&\nu_{i}y_{i}-\frac{1}{2}c_{i}\lambda^{2}-(4\epsilon+\nu_{i})b_{2i}\lambda^{4}+\sum_{j,k}b_{2ijk}(\nu_{j}+\nu_{k}-\nu_{i})y_{j}y_{k}
\end{eqnarray}
which implies the values of couplings $\lambda_{*}$ and $y_{i *}$ at the fixed point of $\mathcal{P}_{2}$,
\begin{eqnarray}{\label{B2}}
-\epsilon+\pi^{2}\lambda^{2}_{*}-4\pi^{4}\sum_{i}c_{i}y_{i*}\mathcal{I}(\nu_{i}, \epsilon) \lambda^{2}_{*}+2\epsilon a_{2}\lambda^{4}_{*}-\sum_{i, j} a_{2ij}(\nu_{i}+\nu_{j})y_{*i}y_{*j}&=&0\nonumber\\
\nu_{i}y_{i *}-\frac{1}{2}c_{i}\lambda^{2}_{*}-(4\epsilon+\nu_{i})b_{2i}\lambda^{4}_{*}+\sum_{j,k}b_{2ijk}(\nu_{j}+\nu_{k}-\nu_{i})y_{*j}y_{*k}&=&0\nonumber\\
\end{eqnarray}

A natural question we have not addressed is under which condition
the approximation up to NNLO  is important and sufficient,
especially in compared with the NLO ones.
For corrections to the second equation in \eqref{B2},
 $y_{i*}\simeq \frac{1}{2}\frac{c_{i}}{\nu_{i}}\lambda^{2}_{*}$ \cite{1203.5129} is always valid
except that the new theory $\mathcal{P}_2$ is beyond the scope of perturbation.
This suggests $y_{i*}<< \lambda^{2}_{*}$ if $c_{i}<<\nu_{i}$,
or equivalently $c_{i}<<1$,
which implies that the effect of $\mathcal{O}(y^{2})$ (even of $\mathcal{O}(\lambda^{2}y)$ ) is smaller
in compared with that of $\mathcal{O}(\lambda^{4})$.
It is necessary to take the order of $\mathcal{O}(\lambda^{4})$ into account and revise those discussions
based on orders up to $\mathcal{O}(\lambda^{2}y)$ but without $\mathcal{O}(\lambda^{4})$,
even though there exists no large hierarchy in the OPE coefficients.
Nevertheless, $y_{i*}>\lambda^{2}$ if $c_{i}\sim \mathcal{O}(1)$.
In this case the corrections arising from $\mathcal{O}(y^{2})$ and $\mathcal{O}(\lambda^{2}y)$ dominate over $\mathcal{O}(\lambda^{4})$.

\section{SCFTs at $\mathcal{O}(\lambda^{4})$}
We perform the perturbative calculations by using the OPEs in appendix A.
The rational is that correlation functions must be independent of $\Lambda$ scale,
which results in the requirement that the coefficients appearing in the same operator that carries  $\Lambda$ factor
must cancel out.
Doing so we obtain,
\begin{eqnarray}{\label{B3}}
a_{2~}&=&16\pi^{4}\left(\frac{c^{2}_{i}}{\nu^{2}_{i}-4\epsilon^{2}}\right)\mathcal{I}(\nu_{i},\epsilon)-\frac{2\pi^{2}}{\epsilon^{2}}\mathcal{T}(\epsilon)\nonumber\\
b_{2i}&=&-\frac{\pi^{2}c_{i}}{2\epsilon(\nu_{i}-2\epsilon)}\left[\mathcal{P}(\nu_{i}, \epsilon)+\mathcal{Q}(\nu_{i}, \epsilon)\right]
\end{eqnarray}
where $\mathcal{I}(\nu_{i},\epsilon)$, $\mathcal{T}(\epsilon)$, $\mathcal{P}(\nu_{i},\epsilon)$ and $\mathcal{Q}(\nu_{i},\epsilon)$
are all dimensionless and smooth functionals as defined in appendix A.

Substituting \eqref{B3} into \eqref{B1}and \eqref{B2} while neglecting the $\mathcal{O}(y^{2})$ effects results in,
\begin{eqnarray}{\label{D6}}
\beta_{\lambda}&=&-\epsilon\lambda+\lambda\left[\pi^{2}\lambda^{2}-4\pi^{4}\sum_{i}c_{i}y_{i}\mathcal{I}(\nu_{i}, \epsilon) \lambda^{2}+2\epsilon a_{2}\lambda^{4}+\cdots\right]\nonumber\\
\beta_{y_{i}}&=&\nu_{i}y_{i}-\frac{1}{2}c_{i}\lambda^{2}-(4\epsilon+\nu_{i})b_{2i}\lambda^{4}+\cdots
\end{eqnarray}
and consequently
\begin{eqnarray}{\label{D7}}
-\epsilon+\pi^{2}\lambda^{2}_{*}-4\pi^{4}\sum_{i}c_{i}y_{i*}\mathcal{I}(\nu_{i}, \epsilon) \lambda^{2}_{*}+2\epsilon a_{2}\lambda^{4}_{*}&=&0\nonumber\\
\nu_{i}y_{i *}-\frac{1}{2}c_{i}\lambda^{2}_{*}-(4\epsilon+\nu_{i})b_{2i}\lambda^{4}_{*}&=&0
\end{eqnarray}
, respectively. The value of $y_{i*}$ is instead of,
\begin{eqnarray}{\label{D8}}
y_{i *}=\frac{c_{i}}{2\nu_{i}}\lambda^{2}_{*}\left[1+\epsilon^{-1}\lambda^{2}_{*}(\mathcal{O}(1)
+\kappa\left(\mathcal{P}(\nu_{i},\epsilon) + \mathcal{Q}(\nu_{i},\epsilon)\right)\right]
\end{eqnarray}
with the coefficient $\kappa$ is strictly of $\mathcal{O}(1)$ no matter how $\nu_{i}$ is relative to $\epsilon$.
So whether the higher-order corrections to $y_{i*}$ in \eqref{D8}  are substantial depend on the finite quantities $\mathcal{P}(\nu_{i},\epsilon)$ and $ \mathcal{Q}(\nu_{i},\epsilon)$.

The $\mathcal{O}(\lambda^{4})$ corrections to $\gamma_{X}(\nu_{i}, \epsilon)$ gives rise to,
 \begin{eqnarray}{\label{D9}}
-\epsilon+\pi^{2}\lambda^{2}_{*}-2\pi^{2}\lambda^{4}_{*}\sum_{i}\frac{1}{{\nu_{i}}}\left[\pi^{2}c^{2}_{i}\left(1-16\frac{\epsilon \nu_{i}}{\nu^{2}_{i}-4\epsilon^{2}}\right)\mathcal{I}(\nu_{i}, \epsilon) -\frac{\nu_{i}}{\epsilon}\left(\frac{3-\epsilon}{2}-2\mathcal{T}(\epsilon)\right)\right]=0\nonumber\\
\end{eqnarray}
Substitute the leading order approximation $\lambda^{2}_{*}\simeq \frac{\epsilon}{\pi^{2}}$
into terms of order $\mathcal{O}(\lambda^{4})$ in \eqref{D9} gives rise to
\begin{eqnarray}{\label{D10}}
\lambda^{2}_{*}\simeq -\frac{\epsilon}{\pi^{2}}+\frac{1}{\pi^{2}}\mathcal{O}\left(\frac{\epsilon^{2} c_{i}^{2}}{\nu_{i}}\right)+\frac{\mathcal{T}(\epsilon)}{\pi^{2}}\mathcal{O}(\epsilon)
\end{eqnarray}
 it is clear to notice that the higher-order corrections can be substantial for determining the fixed point coupling $\lambda_{*}$
when $c_{i}< \nu_{i}$ and even dominate over the order of $\mathcal{O}(\lambda^{2}y_{i})$ when $c_{i}<<\nu_{i}$.
In the region of small $c_{i}$, $c_{i}<<\nu_{i}$,
the $\mathcal{O}(\lambda^{4})$ correction is substantial for determining the fixed point coupling $\lambda_{*}$.

Now we calculate the anomalous dimensions of operators imposed of $L_i$, $X^{\dag}X$ and their mixing,
which can be read from the $\tau$ matrix defined as $\tau\equiv\partial_{(y_{i},\lambda)}\beta_{(y_{i},\lambda)}\mid_{y_{i}*, \lambda_{*}}$.
By using \eqref{D6} we obtain,
 \begin{eqnarray}{\label{D10}}
\tau=\left(\begin{array}{cc}
                  \nu_{i}\delta_{ij} & -c_{i}\lambda_{*}-4\sum_{i}(4\epsilon+\nu_{i})b_{2i}\lambda^{3}_{*} \\
                  -4\pi^{4}\sum_{i}c_{i}\mathcal{I}(\nu_{i},\epsilon) \lambda^{3}_{*} &2\epsilon(1+\frac{5\epsilon^{2}}{\pi^{4}}a_{2})
                  \end{array}\right)
\end{eqnarray}
The deviation of the eigenvalues $\delta$ of this $\tau$ matrix to the case without $\mathcal{O}(\lambda^{4})$ effects
can be more clearly seen after we make a $2\epsilon$ shift in $\tau$,
which is a operation useful for us to directly compare the value of $\delta_{min}$ with \cite{1203.5129},
 \begin{eqnarray}{\label{D11}}
\delta\tau=\left(\begin{array}{cc}
                  (\nu_{i}-2\epsilon)\delta_{ij} & -c_{i}\lambda_{*}-4\sum_{i}(4\epsilon+\nu_{i})b_{2i}\lambda^{3}_{*} \\
                  -4\pi^{4}\sum_{i}c_{i}\mathcal{I}(\nu_{i},\epsilon) \lambda^{3}_{*} &\frac{10\epsilon^{3}}{\pi^{4}}a_{2}
                  \end{array}\right)
\end{eqnarray}

The point is that all the diagonal elements aren't zero,
which remain after a similarity transformation to $\tau$.
So whether there exists such a negative $\delta$ is not obvious anymore.
In general it is quite difficult to obtain the eigenvalues $\delta$ without given the information about relative values of $\nu_{i}$ and $\epsilon$.
We divide this task into a few cases.
The first ,  also trivial case is  $\nu_{i}<<\epsilon<<1$,
in which there are already some $L_i$ with dimension smaller than $2\Delta_X$.
The other cases $\epsilon << \nu_{i}<<1$ and $\epsilon \sim \nu_{i}<<1$ are of more interest to us.

\subsection{$\epsilon << \nu_{i}<<1$}
Now we address the simplification for the functionals as defined in appendix A in the region $\epsilon << \nu_{i}<<1$.
Each integral variable $X^{+}_i$ in these functionals are evaluated in the region $|X^{+}_i|>\frac{1}{\Lambda}$,
with $\Lambda$ the cut-off scale introduced in \eqref{A4},
and integral over Grassmann variables is equivalent to performing derivative over them.
For functional $\mathcal{I}(\nu_{i},\epsilon)$ \eqref{I},
performing the integral gives us,
 \begin{eqnarray}{\label{D12}}
\mathcal{I}(\nu_{i},\epsilon)|_{\nu_{i}<<1,~\epsilon<<1}\simeq 1+\mathcal{O}(\epsilon, \nu_{i})
\end{eqnarray}

Similar operation can be applied to $\mathcal{P}(\nu_{i}, \epsilon)$ functional,
which explicitly reads,
 \begin{eqnarray}{\label{D13}}
\mathcal{P}(\nu_{i},\epsilon)|_{\nu_{i}<<1,~\epsilon<<1}&\simeq & \frac{2\epsilon-\nu_{i}}{2\epsilon+\nu_{i}}+\mathcal{O}(\epsilon, \nu_{i})\simeq -1+\mathcal{O}(\epsilon, \nu_{i})
\end{eqnarray}
after setting $\nu_{i}=0$ and replacing $2\epsilon\rightarrow 2\epsilon+\nu_{i}$.
The functionals $\mathcal{Q}(\nu_{i}, \epsilon)$ and $\mathcal{T}(\epsilon)$ are three-dimensional integrals,
thus more involved than $\mathcal{I}(\nu_{i},\epsilon)$ and $\mathcal{P}(\nu_{i},\epsilon)$.
For this case one can integrate over one variable,
then follow the similar operation for the two-dimensional integral.
At leading order, we find
 \begin{eqnarray}{\label{D14}}
\mathcal{Q}(\nu_{i},\epsilon)|_{\nu_{i}<<1,~\epsilon<<1}&\simeq & \left(\frac{2\epsilon-\nu_{i}}{2\epsilon+\nu_{i}}\right) \epsilon\Gamma(2\epsilon)+\cdots =-1+\mathcal{O}(\epsilon, \nu_{i})\nonumber\\
\mathcal{T}(\nu_{i},\epsilon)|_{\nu_{i}<<1,~\epsilon<<1}&\simeq & + \epsilon^{2}\left[\Gamma(2\epsilon)+\cdots\right]=+\mathcal{O}(\epsilon)+\cdots
\end{eqnarray}
where we have ignored the higher-order terms.
The coefficients at the leading order,
related to the complicated Hypergeoemtric function $~_{2}F_{1}(1, m-2\epsilon, 1+2\epsilon, -1)$ (with integer $m$),
are finite and not shown explicitly.
We will see the approximations \eqref{D12}- \eqref{D14}
are sufficient to illustrate the modification to anomalous dimensions of $\mathcal{S}_i$.

Substitute \eqref{D12}- \eqref{D14} into \eqref{B3}, one can substantially simplify $a_2$ and $b_{2i}$.
Doing so, we obtain the leading-order approximation to the matrix $\delta\tau$ in \eqref{D11} under the limit $\epsilon << \nu_{i}<<1$,
 \begin{eqnarray}{\label{D15}}
\delta\tau &=&\left(\begin{array}{cc}
                  \nu_{i}\delta_{ij} & -c_{i}\lambda_{*}\left[1+\frac{1}{\epsilon}(3+4\pi^{2})\right] \\
                  -4\pi^{4}\sum_{i}c_{i} \lambda^{3}_{*} & -\frac{20\epsilon}{\pi^{2}}\mathcal{T}(\epsilon)
                  \end{array}\right)
\end{eqnarray}
Put the values of couplings at NLO back into \eqref{D15},
the characteristic equation of $\delta$ is found to be,
 \begin{eqnarray}{\label{D16}}
\left(\nu_{i}\delta_{ij}-\delta\right)\left(-\frac{20\epsilon}{\pi^{2}}\mathcal{T}(\epsilon) -\delta\right)-4\pi^{2}(3+4\pi^{2})c^{2}_{i}\epsilon=0
\end{eqnarray}
Together with the small $c_i<< \nu_{i}$ condition assumed through out this section,
we notice that the last constant term in \eqref{D16} is actually small compared with $\nu_{i}\epsilon$ if $c_{i}$ is below the critical value $c_{i*}\simeq\sqrt{\nu_{i}\epsilon}$,
which implies that the minimal value of $\delta$ is of order $\epsilon^{2}$,
 \begin{eqnarray}{\label{D17}}
\delta_{min}\simeq - \frac{20\epsilon}{\pi^{2}}\mathcal{T}(\epsilon)+\mathcal{O}(\epsilon^{3})  <0
\end{eqnarray}
One thing happens when $c_{i}$ is above the critical value $c_{i*}$.
The last term dominate conversely, which modified the \eqref{D16} as,
\begin{eqnarray}{\label{D18}}
\delta_{min}\simeq -\pi^{2}(3+4\pi^{2})\frac{c^{2}_{i}\epsilon}{\nu_{i}} \simeq -\pi^{2}(3+4\pi^{2})\left(\frac{c_{i}}{c_{i*}}\right)^{2}\epsilon^{2}< 0
\end{eqnarray}

\subsection{$\nu_{i}\sim \epsilon<<1$}
Since it is quite natural to expect that $\nu_{i}$ is of $\mathcal{O}(\epsilon)$ or higher powers of $\epsilon$  in perturbative CFT,
a number of $\mathcal{P}_{2}$ theories can be covered in this limit.
Now we address the question that whether the statement  in the previous discussion
can be generalized to this particular situation.
At first,  $a_2$ and $b_{2i}$ take the approximation
\footnote{Note that $a_2$ has a pole at $\nu_{i}=2\epsilon$. Here, we assume $\nu_i$ is not equal to $2\epsilon$ for simplification.},
\begin{eqnarray}{\label{D19}}
a_{2}&=&-\frac{2\pi^{2}}{\epsilon^{2}}\mathcal{T}(\epsilon)\nonumber\\
b_{2i}&=&-\frac{3\pi^{2}c_{i}}{4\epsilon(\nu_{i}+2\epsilon)}
\end{eqnarray}
Substitute these values into \eqref{D11}, we obtain
\begin{eqnarray}{\label{D20}}
\delta\tau =\left(\begin{array}{cc}
                  (\nu_{i}-2\epsilon)\delta_{ij} & -c_{i}\lambda_{*}\left[1+\mathcal{O}(\epsilon^{-1}\lambda^{2})\right] \\
                  -4\pi^{4}\sum_{i}c_{i} \lambda^{3}_{*} & -\frac{20\epsilon}{\pi^{2}}\mathcal{T}(\epsilon)
                  \end{array}\right)
=\left(\begin{array}{cc}
                  (\nu_{i}-2\epsilon)\delta_{ij} & \mathcal{O}(c_{i}^{\frac{1}{2}}) \\
                  \mathcal{O}(c_{i}\epsilon^{\frac{3}{2}}) & -\frac{20\epsilon}{\pi^{2}}\mathcal{T}(\epsilon)
                  \end{array}\right)
\end{eqnarray}
Drop the off-diagonal elements in above matrix by using the relation $c_{i}<< \nu_{i}\sim\epsilon$,
we arrive at the conclusion that the statement is also true in the region.

 \em{In summary, if   $c_{i}<<\nu_{i}<<1$  is indeed produced given a $\mathcal{P}_{2}$
theory, then we can conclude that the bound on the anomalous dimension of $\mathcal{S}_i$ as conjectured in the literature is still valid at NNLO ,
no matter the relative values of $\epsilon$ and $\nu_{i}$.}\em
Therefore, the validity of this conjecture is directly transferred to examine these conditions in $\mathcal{P}_{2}$ theory
\footnote{We want to remind the reader that naively this statement can not be directly applied to BZ theory with large $N$ limit.
However, in BZ theory $\nu_{i}\simeq\mathcal{O}(\epsilon^{2})$ \cite{1203.5129},
which actually suggests some of anomalous dimension of $L_i$ is already smaller than that of $X$.
This statement is trivially satisfied in this situation. }.

\section{SCFTs at $\mathcal{O}(y^{2})$}
The $\mathcal{O}(y^{2})$ corrections dominate over $\mathcal{O}(\lambda^{4})$ when $c_{i}>>\nu_{i}$.
The investigation of bounds on $c_i$ can be found in \cite{1109.5176, C1}.
Instead of calculating the wave-function renormalization and beta function as in appendix A,
on must consider $L_i$ operators.
But this task can not be precisely achieved without knowing the explict form of $L_i$ (for example $L_i$ are composite operators).
The $\mathcal{O}(y^{2})$ effects can only be analyzed either in a specific $\mathcal{P}_{1}$ theory or in certain approximations.

\subsection{BZ Theory As an illustration}
One might wonder which  $\mathcal{P}_{1}$ theory can provide such kind of condition.
Actually,  given a special choice of the flavor number $N_f$ and rank of gauge group $N_c$,
the BZ theory  \cite{BZ} could be a simple realization.
It is classified in \cite{1203.5129} that $L=bTr(Q^{\dag}Q+\tilde{Q}^{\dag}\tilde{Q})$ in the BZ theory,
with $Q_i$ being the chiral matter superfields.
Under the large $N$ limit with $\frac{N_{f}}{3N_{c}}=1+\epsilon$ and normalizations taken in Ref \cite{1203.5129} ,
it is found that $c_{L}=\sqrt{\frac{2}{N_{f}N_{c}}}$ and $\nu_{L}\simeq 3\epsilon^{2}$.
Impose the constraint $c_{i}<<\nu_{i}$, we find $\epsilon^{2}<<\frac{1}{N_c}$.
Take the perturbative condition $y\simeq \frac{c_{i}}{\nu_{i}}\lambda^{2}\simeq \frac{c_{i}\epsilon}{\nu_{i}}$ into account ,
we obtain  $\epsilon>>\frac{1}{N_c}$ for consistency.
So if $\epsilon$ which can be considered as an input parameter is left to be
in the narrow window
\begin{eqnarray}{\label{H1}}
\frac{1}{N_c}<<\epsilon<< \frac{1}{\sqrt{N_c}}
\end{eqnarray}
then higher-order corrections in this BZ theory arising from $\mathcal{O}(y^{2})$
indeed dominate over $\mathcal{O}(\lambda^{4})$.

To estimate the $\mathcal{O}(y^{2})$  corrections to the matrix of anomalous dimensions at NLO \cite{1203.5129},
\begin{eqnarray}{\label{H2}}
\tau =\left(\begin{array}{cc}
                  \nu_{L}\simeq 3\epsilon^{2} & -\frac{3\epsilon^{2}}{N^{2}_{c}}\\
                  -\frac{4}{3}\epsilon  & 2\epsilon
                  \end{array}\right)
\end{eqnarray}
one must consider the higher-order terms in the anomalous dimensions of $Q$ and $X$,
especially those unsuppressed by $1/N$.
From \cite{9311340} (see also \cite{1203.5129}) we obtain,
\begin{eqnarray}{\label{H3}}
\delta\gamma_{Q}(\hat{g},\lambda)=\frac{2-\epsilon}{1+\epsilon}\hat{g}^{2}+\mathcal{O}(\hat{g}^{2}/N^{2}_{c}),~~~~~~~~~
\delta\gamma_{X}(\hat{g},\lambda)\simeq\frac{\hat{g}\hat{\lambda}}{N_{c}^{2}}++\mathcal{O}(\hat{g}\hat{\lambda}/N^{2}_{c})
\end{eqnarray}
where $\hat{g}=\frac{N_{c}g^{2}}{16\pi^{2}}$.
Substituting \eqref{H3} into the $\tau$ matrix leads to correction to \eqref{H2},
\begin{eqnarray}{\label{H4}}
\delta\tau =\left(\begin{array}{cc}
                  -12\hat{g}^{3}_{*} & 0 \\
                  \frac{16}{3}\epsilon^{2}  & \frac{4\epsilon^{2}}{N^{2}_{c}}
                  \end{array}\right)=
\left(\begin{array}{cc}
                  \mathcal{O}(\epsilon^{3}) & ~0 \\
                  \mathcal{O}(\epsilon^{2}) & ~\mathcal{O}(\epsilon^{3})\sim \mathcal{O}(\epsilon^{4})
                  \end{array}\right)<< \tau
\end{eqnarray}
by using the constraint \eqref{H1}.
Unlike the situation in the previous section,
each matrix element is smaller compared with those at NLO in this case.
This suggests that the ability to affect the sign of $\delta_{min}$ coming from
 $\mathcal{O}(y^{2})$  is weaker than $\mathcal{O}(\lambda^{4})$.

\subsection{Analysis of OPE}
The simple example of BZ theory in the previous discussion
provides us an intuition that the $\mathcal{O}(y^{2})$ corrections are probably negligible under the assumptions taken by us in the setup.
Now we address this issue by analyzing the OPEs in this case.
The estimate of  $\mathcal{O}(y^{2})$ effects involved the calculations of coefficients $a_{2ij}$ and $b_{2ijk}$.
The possible combinations that contribute to coefficient $b_{2ijk}$ are \em { null}\em ~due to the fact
that all of the coefficients at $\mathcal{O}(y)$ vanish in \eqref{D4} and \eqref{D5} .
For $a_{2ij}$, by using the results in  \eqref{D4} and \eqref{D5} all the combinations of operators do not contribute,
which gives us
\begin{eqnarray}{\label{D21}}
a_{2ij}=0,~~~~~and~~~~b_{2ijk}=0
\end{eqnarray}

\em{In summary, the NNLO corrections due to $\mathcal{O}(y^{2})$ are actually null.
~The statement in the previous section holds also in the region of $c_{i}>>\nu_{i}$
(but still on the realm of perturbative field theory)}\em.

What about the higher-order terms involved $y_i$ couplings.
The vanishing contributions both at NLO and NNLO indicates
that the contributions arising from $y_i$ beyond NLO do not exist, i.e,
the coefficients in powers of $y^{n}_{i}\lambda^{m}$ ($n=2,3, \cdots $, $m=0, 1, \cdots$ ) are \em{exactly}\em~ zero.
In general, these operators are related to the following OPEs,
\begin{eqnarray}{\label{D22}}
&~&X^{\dag}(z^{-}_{1})L_{i}(x_{2},\theta_{2},\bar{\theta}_{2}) \nonumber\\
&~&X(z^{+}_{1})L_{i}(x_{2},\theta_{2},\bar{\theta}_{2})\\
&~&L_{i}(x_{1},\theta_{1},\bar{\theta}_{1})L_{j}(x_{2},\theta_{2},\bar{\theta}_{2})  \nonumber
\end{eqnarray}

To determine the OPEs in \eqref{D22},
we use a  crucial observation in our setup.
At first, the primary operators $\mathcal{S}_i$ are composed of primary operators $L_i$ and $X^{\dag}X$
because of the interaction mediated by $\lambda$.
This implies that $\mathcal{S}_i$ can be generally expressed as\footnote{We understand this expression is not exact from the viewpoint of superconformal symmetries, but it indeed captures the main property of scaling dimension relevance,
which is the central concern of this note. Also note that operators composed of (super)derivative over operators on the RHS of \eqref{D23} are not permitted.},
\begin{eqnarray}{\label{D23}}
\mathcal{S}_i (x)=\frac{\cos\alpha_{i}}{\mid x\mid^{\tilde{\Delta}_{i}-\Delta_{i}}}L_{i}-\frac{\sin\beta_{i}}{\mid x\mid^{\tilde{\Delta}_{i}-\Delta_{X^{\dag}X}}}X^{\dag}X(x) +\cdots
\end{eqnarray}
Angle $\alpha_i$ and $\beta_i$ are introduced to represent the mixings.
Here we refer $\tilde{\Delta}_{i}$ to the scaling dimension of $\mathcal{S}_i$.
What are ignored in \eqref{D23} are irrelevant for our purpose.
Define the $d_i$ as the OPE coefficient in three-point correlator :
\begin{eqnarray}{\label{D24}}
<X^{\dag}(z^{-}_{2}, \bar{\theta}_{2})X(z^{+}_{1}, \theta_{1})\mathcal{S}_i (x_{3}, \theta_{3}, \bar{\theta}_{3})>
=\frac{d_{i}}{(X^{+}_{21})^{2\Delta_{X}-\tilde{\Delta}_{i}}(X^{+}_{23})^{\tilde{\Delta}_{i}}(X^{+}_{31})^{\tilde{\Delta}_{i}}}
\end{eqnarray}
We can subtract the OPEs in \eqref{D22} by the OPEs of $\mathcal{S}_i$s.
From \eqref{D24} we obtain the two-point OPEs:
\begin{eqnarray}{\label{D25}}
\mathcal{S}(x_{3},\theta_{3},\bar{\theta}_{3})X^{\dag}(z^{-}_{2}, \bar{\theta}_{2})&\rightarrow&\frac{d_{i}}{(X^{+}_{23})^{\tilde{\Delta}_{i}}}X^{\dag}(z^{-}_{2}, \bar{\theta}_{2})+\cdots\nonumber\\
\mathcal{S}(x_{3},\theta_{3},\bar{\theta}_{3})X(z^{+}_{1}, \theta_{1})&\rightarrow&\frac{d_{i}}{(X^{+}_{31})^{\tilde{\Delta}_{i}}}X(z^{+}_{1}, \theta_{1})+\cdots
\end{eqnarray}

Now we derive the OPEs  in \eqref{D22}.
 From \eqref{D23} we obtain,
\begin{eqnarray}{\label{D26}}
L_{i}\simeq \frac{\cos\alpha_{i}}{\mid x \mid^{\Delta_{i}-\tilde{\Delta}_{i}}}\mathcal{S}_{i}(x)+\frac{\sin\beta_{i}}{\mid x\mid^{\Delta_{i}-2}}X^{\dag}X(x)+\cdots
\end{eqnarray}
Consequently, the OPEs \eqref{D22} can be derived in terms of \eqref{D26}, \eqref{D24} and \eqref{D25},
\begin{eqnarray}{\label{D27}}
L_{i}(x_{1}, \theta_{1},\bar{\theta}_{1})L_{j}(x_{2}, \theta_{2},\bar{\theta}_{2})
\rightarrow \frac{\sin\beta_{i}\sin\beta_{j}}{\mid x_{1} \mid^{\Delta_{i}-2}\mid x_{2}\mid^{\Delta_{j}-2}(X_{21}^{+})^{2}}X^{\dag}X(x_{1},\theta_{1},\bar{\theta}_{2})+\cdots
\end{eqnarray}
and
\begin{eqnarray}{\label{D28}}
&~&X^{\dag}X(x_{1},\theta_{1}, \bar{\theta}_{1})L_{i}(x_{2},\theta_{2},\bar{\theta}_{2}) L_{i}(x_{3},\theta_{3},\bar{\theta}_{3})\rightarrow
\nonumber\\
&~&\frac{1}{(X^{+}_{12})^{\tilde{\Delta}_{i}}(X^{+}_{31})^{\tilde{\Delta}_{i}}\mid x_{2}\mid^{\Delta_{i}-\tilde{\Delta}_{i}} \mid x_{3}\mid ^{\Delta_{j}-\tilde{\Delta}_{i}}}\left[\cos\alpha_{i}\cos\alpha_{j}d_{i}d_{j}X^{\dag}X(x_{1}, \theta_{1},\bar{\theta}_{2})
+\cdots \right]\nonumber\\
\end{eqnarray}
where $....$  in the second line in \eqref{D28} refer to similar structure of $X^{\dag}X$.

Consider the coefficient $a_{3ij}$ that appears in $a_{3ij}y_{i}y_{j}\lambda^{2}$ as an example at the next-to-NNLO.
The combinations arising from multiple $X^{\dag}X$ themselves do not contribute, with only those possibilities in \eqref{D22} left.
Substitute \eqref{D28} and \eqref{D27} into the operators that contribute to $a_{3ij}y_{i}y_{j}\lambda^{2}$ ,
we find that both of them vanish due to the residual Grassmann integrals.
We conclude that the claim on null contribution coming from $y_{i}$ coupling beyond NLO still holds.

\section{Conclusions}
In this note we study the effects of NNLO corrections on the conjecture that $\delta_{min}<0$,
 in the context of perturbative CFT.
As we have emphasized,
despite smaller than NLO ones,
the NNLO corrections are important and even substantial in some circumstances.
In particular, the modifications to the vanishing matrix elements of anomalous dimension at NLO can directly affect the sign of $\delta_{min}$,
although they don't substantially modify the values of fixed points couplings $\lambda_{*}$ and $y_{i *}$.

The main results include:
\begin{enumerate}
\item In the region of  $c_{i}<<\nu_{i}<<1$  in a $\mathcal{P}_{2}$ theory as defined in the introduction,
the bound on the anomalous dimension of $\mathcal{S}_i$ as conjectured in the literature is still valid at NNLO,
no matter the relative values of $\epsilon$ and $\nu_{i}$.
\item In the region of $c_{i}>>\nu_{i}$ the NNLO corrections due to $\mathcal{O}(y^{2})$ effects are actually null.
the conjecture still holds.
\item  The null contribution arising from $y_i$ couplings beyond NLO \em{exactly}\em~remains.
\end{enumerate}
There are a few points that deserve further investigation.
For instance, one can examine the conjecture in background of strongly coupled SCFTs via method of ADS/CFT.
Throughout this note, we have not addressed the possibility that there are  residual global symmetries after imposing the deformation,
it would be also interesting to discuss this issue in the further.

~~~~~~~~~~~~~~~~~~~~~~~~~~~~~~~~~~~~~~~~
$\bf{Acknowledgement}$\\
We would like to thank Tianjun Li, Jia-Hui Huang, Wei-Shui Xu for communications, 
and the referee for valuable suggestions.
This work is supported in part by the Fundamental Research Funds for the Central Universities with Grant No. CDJZR11300001.\\

\appendix
\section{OPEs and $\mathcal{O}(\lambda^{4})$ Effects}
In superspace , the  two-point functions for $\mathcal{O}\mathcal{O}^{\dag}$, $XX^{\dag}$
and three-point function for $L\mathcal{O}\mathcal{O}^{\dag}$ are given by \cite{Osborn, 1203.5129},
\begin{eqnarray}{\label{A1}}
<\mathcal{O}(z^{+}_{1}, \theta_{1})\mathcal{O}^{\dag}(z^{-}_{2}, \bar{\theta}_{2})>&=&\frac{1}{(X^{+}_{21})^{2(2-\epsilon)}}\nonumber\\
<X(z^{+}_{1}, \theta_{1})X^{\dag}(z^{-}_{2}, \bar{\theta}_{2})>&=&\frac{1}{(X^{+}_{21})^{2}}\\
<\mathcal{O}(z^{+}_{1}, \theta_{1})\mathcal{O}^{\dag}(z^{-}_{2}, \bar{\theta}_{2})L(x_{3},\theta_{3},\bar{\theta}_{3})>&=&\frac{c_{i}}{(X^{+}_{21})^{2-2\epsilon-\nu_{i}}(X^{+}_{23})^{2+\nu_{i}}(X^{+}_{31})^{2+\nu_{i}}}\nonumber
\end{eqnarray}
where $X^{+}_{ij}=z^{-}_{i}-z^{+}_{j}+2i\theta_{j}\sigma\bar{\theta}_{i}$ is a supertranslation invariant interval.
Here $z^{\pm}=x\pm i\theta\sigma\bar{\theta}$.
We also need the following superspace OPEs that can be derived from \eqref{A1},
\begin{eqnarray}{\label{A2}}
\mathcal{O}^{\dag}(z^{-}_{2}, \bar{\theta}_{2})\mathcal{O}(z^{+}_{1}, \theta_{1})
&\rightarrow&\frac{1}{(X^{+}_{21})^{2(2-\epsilon)}}+\frac{c_{i}}{(X^{+}_{21})^{2-2\epsilon-\nu_{i}}}L_{i}+\cdots \nonumber\\
X^{\dag}(z^{-}_{2}, \bar{\theta}_{2})X(z^{+}_{1}, \theta_{1})&=&\frac{1}{(X^{+}_{21})^{2}}+X^{\dag}X(x_{1},\theta_{1}, \bar{\theta}_{2})+\cdots
\end{eqnarray}
and
\begin{eqnarray}{\label{A3}}
L(x_{3},\theta_{3},\bar{\theta}_{3})\mathcal{O}^{\dag}(z^{-}_{2}, \bar{\theta}_{2})&=&\frac{c_{i}}{(X^{+}_{23})^{2+\nu_{i}}}\mathcal{O}^{\dag}(z^{-}_{2}, \bar{\theta}_{2})+\cdots\nonumber\\
L(x_{3},\theta_{3},\bar{\theta}_{3})\mathcal{O}(z^{+}_{1}, \theta_{1})&=&\frac{c_{i}}{(X^{+}_{31})^{2+\nu_{i}}}\mathcal{O}(z^{+}_{1}, \theta_{1})+\cdots
\end{eqnarray}
The terms ignored in \eqref{A2} and \eqref{A3} are superconformal descendant,
 which are irrelevant for our calculations of beta function.

The terms involved in $\mathcal{O}(\lambda^{4})$ wave-function renormalization can be read from \eqref{D1} and \eqref{D3},
\begin{eqnarray}{\label{A4}}
 \frac{1}{4\pi^{2}}\int  d^{4}xd^{4}\theta(1+a_{1}\lambda^{2}+a_{2}\lambda^{4}+\cdots)X^{\dag}X
+ \int  d^{4}xd^{4}\theta (y_{i}+b_{1i} \lambda^{2}+b_{2i}\lambda^{4}+\cdots )\lambda^{-\nu_{i}} L_{i}\nonumber\\
+\frac{\lambda}{2\pi}\left(\int d^{4}z^{+}_{2}d\theta^{2}_{2}\Lambda^{\epsilon}~\mathcal{O}X(z_{2}^{+}, \theta_{2})
+\int d^{4}z^{-}_{1} d\bar{\theta}^{2}_{1} \Lambda^{\epsilon}~\mathcal{O}^{\dag}X^{\dag}(z_{1}^{-}, \bar{\theta}_{1})\right)\nonumber\\
\end{eqnarray}
Evaluating \eqref{A4} we obtain the counter terms of order $\mathcal{O}(\lambda^{4})$,
\begin{eqnarray}{\label{A5}}
&\lambda^{4}&\left[ \frac{a_{2}}{4\pi^{2}}\int  d^{4}x d^{4}\theta X^{\dag}X (x, \theta, \bar{\theta})+
b_{2i}\lambda^{-\nu_{i}} \int  d^{4}xd^{4}\theta L_{i} \right.\nonumber\\
&+&\left. \frac{b_{2i}}{4\pi^{2}}\Lambda^{-\nu_{i}}\int  d^{4}x_{1}d^{4}x_{2}d^{4}\theta_{1}d^{4}\theta_{2}~X^{\dag}X(x_{2},\theta_{2}, \bar{\theta}_{2})L_{i}(x_{1},\theta_{1}, \bar{\theta}_{1})\right.\nonumber\\
&+&\left.\frac{1}{(4\pi^{2})^{2}}\int  d^{4}x_{1}d^{4}x_{2}d^{4}\theta_{1}d^{4}\theta_{2}~X^{\dag}X(x_{2},\theta_{2}, \bar{\theta}_{2})~
X^{\dag}X(x_{1},\theta_{1}, \bar{\theta}_{1})\right.\nonumber\\
&+&\left. \frac{a_{1}}{(4\pi^{2})^{2}}\Lambda^{2\epsilon} \int d^{4}z^{-}_{1}d^{4}z^{+}_{2}d^{4}x_{3}d^{2}\bar{\theta}_{1}d^{2}\theta_{2}d^{4}\theta_{3}~
\mathcal{O}^{\dag}X^{\dag}(z_{1}^{-}, \bar{\theta}_{1}) ~\mathcal{O}X(z_{2}^{+}, \theta_{2})~X^{\dag}X(x_{3},\theta_{3}, \bar{\theta}_{3})\right.\nonumber\\
&+&\left. \frac{b_{1i}}{4\pi^{2}}\Lambda^{2\epsilon-\nu_{i}}~ \int d^{4}z^{-}_{1}d^{4}z^{+}_{2}d^{4}x_{3}d^{2}\bar{\theta}^{2}_{1}d^{2}\theta_{2}d^{4}\theta_{3}~
\mathcal{O}^{\dag}X^{\dag}(z_{1}^{-}, \bar{\theta}_{1}) ~\mathcal{O}X(z_{2}^{+}, \theta_{2})~
L_{i}(x_{3},\theta_{3}, \bar{\theta}_{3})\right.\nonumber\\
&+&\left. \frac{1}{(4\pi^{2})^{2}}\Lambda^{4\epsilon}~
\int d^{4}z^{-}_{1}d^{4}z^{+}_{2}d^{4}z^{-}_{3}d^{4}z^{+}_{4}
d^{2}\bar{\theta}_{1}d^{2}\theta_{2}d^{2}\bar{\theta}_{3}d^{2}\theta_{4}~
\mathcal{O}^{\dag}X^{\dag}(z_{1}^{-}, \bar{\theta}_{1}) ~\mathcal{O}X(z_{2}^{+}, \theta_{2})\right.\nonumber\\
&\times& \left.\mathcal{O}^{\dag}X^{\dag}(z_{3}^{-}, \bar{\theta}_{3}) ~\mathcal{O}X(z_{4}^{+}, \theta_{4})
\right]
\end{eqnarray}
which gives us,
\begin{eqnarray}{\label{A6}}
-b_{2i}&=&\left[\frac{a_{1}}{(4\pi^{2})^{2}}\Lambda^{2\epsilon+\nu_{i}} \int d^{4}z^{-}_{1}d^{4}x_{3}d^{4}\theta_{3}~
\frac{c_{i}}{(X^{+}_{32})^{2}(X^{+}_{13})^{2}(X^{+}_{12})^{2-2\epsilon-\nu_{i}}}\right.\nonumber\\
&+&\left.  \frac{b_{1i}}{4\pi^{2}}\Lambda^{2\epsilon}
~ \int d^{4}z^{-}_{1}d^{4}z^{+}_{2}d^{2}\bar{\theta}_{1}d^{2}\theta_{2}~\frac{1}{(X^{+}_{12})^{6-2\epsilon}}\right.\nonumber\\
&+&4\times \left.\frac{c_{i}}{(4\pi^{2})^{2}}\Lambda^{4\epsilon+\nu_{i}}
 \int \frac{d^{4}z^{-}_{1}d^{4}z^{+}_{2}d^{4}z^{-}_{3}d^{2}\theta_{2}d^{2}\bar{\theta}_{3}}
{(X^{+}_{12})^{2}(X^{+}_{34})^{2}(X^{+}_{14})^{2-2\epsilon-\nu_{i}}(X^{+}_{32})^{2(2-\epsilon)}}
 \right]
\end{eqnarray}
from the last three terms in \eqref{A5} and OPEs given in \eqref{A2}.
The factor 4 in the last line in \eqref{A6} counts the four symmetric permutations.
Performing the intergral of the first line in \eqref{A6} gives,
\begin{eqnarray}{\label{A7}}
\frac{a_{1}c_{i}}{(4\pi^{2})^{2}}\Lambda^{2\epsilon+\nu_{i}} \int
\frac{d^{4}X^{+}_{13}d^{4}X_{32}d^{4}\theta_{32}}{(X^{+}_{32})^{2}(X^{+}_{13})^{2}(X^{+}_{12})^{2-2\epsilon-\nu_{i}}}
\equiv \frac{1}{2(\nu_{i}-2\epsilon)}a_{1}c_{i} \mathcal{P}(\nu_{i}, \epsilon)
\end{eqnarray}
with
\begin{eqnarray}{\label{A8}}
\mathcal{P}(\nu_{i}, \epsilon)&=&\frac{(\nu_{i}-2\epsilon)}{8\pi^{4}}\Lambda^{2\epsilon+\nu_{i}} \int
\frac{d^{4}X^{+}_{13}d^{4}X_{32}d^{4}\theta_{32}}{(X^{+}_{32})^{2}(X^{+}_{13})^{2}(X^{+}_{12})^{2-2\epsilon-\nu_{i}}}\nonumber\\
&=&\frac{(\nu_{i}-2\epsilon)(3-2\epsilon-\nu_{i})(2-2\epsilon-\nu_{i})}{8\pi^{4}}\Lambda^{2\epsilon+\nu_{i}} \int
\frac{d^{4}X^{+}_{13}d^{4}X^{+}_{32}}{(X^{+}_{32})^{2}(X^{+}_{13})^{2}(X^{+}_{13}+X^{+}_{32})^{4-2\epsilon-\nu_{i}}}\nonumber\\
\end{eqnarray}
where we have changed the integration variables $z^{-}_{1}\rightarrow X^{+}_{13}$, $x_{3}\rightarrow X^{+}_{32}$
, $\theta_{3}\rightarrow \theta_{32}$  and $\bar{\theta}_{3}\rightarrow \bar{\theta}_{32}$, and use the equality
 $X^{+}_{12}=X^{+}_{13}+X^{+}_{32}+2i\theta_{32}\sigma\bar{\theta}_{32}$ .
The second integral in \eqref{A6} is equal to,
\begin{eqnarray}{\label{A9}}
 \frac{b_{1i}}{4\pi^{2}}\Lambda^{2\epsilon}
~ \int \frac{d^{4}X^{+}_{12}d^{4}X^{+}_{23}d^{2}\bar{\theta}_{12}d^{2}\theta_{23}}{(X^{+}_{12})^{6-2\epsilon}}
=0
\end{eqnarray}
after we are free to change the integral variables $z^{-}_{1}\rightarrow X^{+}_{12}$, $z^{+}_{2}\rightarrow X^{+}_{23}$,
$\bar{\theta}_{1} \rightarrow \bar{\theta}_{12}$ and $\theta_{2} \rightarrow\theta_{23}$.
The last integral in \eqref{A6}  can be reexpressed as,
\begin{eqnarray}{\label{A10}}
4\times \frac{c_{i}}{(4\pi^{2})^{2}}\Lambda^{4\epsilon+\nu_{i}}
 \int \frac{d^{4}z^{-}_{1}d^{4}z^{+}_{2}d^{4}z^{-}_{3}d^{2}\theta_{2}d^{2}\bar{\theta}_{3}}
{(X^{+}_{12})^{2}(X^{+}_{34})^{2}(X^{+}_{14})^{2-2\epsilon-\nu_{i}}(X^{+}_{32})^{2(2-\epsilon)}}
\equiv \frac{\pi^{2} c_{i}}{2\epsilon(\nu_{i}-2\epsilon)} \mathcal{Q}(\nu_{i}, \epsilon)
\end{eqnarray}
with
\begin{eqnarray}{\label{A11}}
\mathcal{Q}(\nu_{i}, \epsilon)&=&\frac{\epsilon(\nu_{i}-2\epsilon)}{2\pi^{6}}\Lambda^{4\epsilon+\nu_{i}}
 \int \frac{d^{4}z^{-}_{1}d^{4}z^{+}_{2}d^{4}z^{-}_{3}d^{2}\theta_{2}d^{2}\bar{\theta}_{3}}
{(X^{+}_{12})^{2}(X^{+}_{34})^{2}(X^{+}_{14})^{2-2\epsilon-\nu_{i}}(X^{+}_{32})^{2(2-\epsilon)}}\nonumber\\
&=&\frac{\epsilon(\nu_{i}-2\epsilon)}{2\pi^{6}}\Lambda^{4\epsilon+\nu_{i}}
 \int \frac{d^{4}X^{+}_{12}d^{4}X^{+}_{32}d^{4}X^{+}_{34}d^{2}\theta_{42}d^{2}\bar{\theta}_{13}}
{(X^{+}_{12})^{2}(X^{+}_{34})^{2}(X^{+}_{12}-X^{+}_{32}+X^{+}_{34}+2i\theta_{42}\sigma\bar{\theta}_{13})^{2-2\epsilon-\nu_{i}}(X^{+}_{32})^{2(2-\epsilon)}}\nonumber\\
&=&\frac{\epsilon(\nu_{i}-2\epsilon)(3-2\epsilon-\nu_{i})(2-2\epsilon-\nu_{i})}{2\pi^{6}}\Lambda^{4\epsilon+\nu_{i}}
\nonumber\\
&\times& \int \frac{d^{4}X^{+}_{12}d^{4}X^{+}_{32}d^{4}X^{+}_{34}}
{(X^{+}_{12})^{2}(X^{+}_{34})^{2}(X^{+}_{12}-X^{+}_{32}+X^{+}_{34})^{4-2\epsilon-\nu_{i}}(X^{+}_{32})^{2(2-\epsilon)}}
\end{eqnarray}
after we change the integral variables  $z^{-}_{1}\rightarrow X^{+}_{12}$,  $z^{+}_{2}\rightarrow -X^{+}_{32}$,
$z^{-}_{3}\rightarrow X^{+}_{34}$,
$\theta_{2}\rightarrow -\theta_{42}$ and  $\bar{\theta}_{3}\rightarrow-\bar{\theta}_{13} $ and use the equality
$X^{+}_{14}=X^{+}_{12}-X^{+}_{32}+X^{+}_{34}+2i\theta_{42}\sigma\bar{\theta}_{13}$.

Collect the results in \eqref{A10}, \eqref{A9}and  \eqref{A7}, we have the final result about $b_{2i}$,
\begin{eqnarray}{\label{A12}}
b_{2i}=-\frac{\pi^{2}c_{i}}{2\epsilon(\nu_{i}-2\epsilon)}\left[\mathcal{P}(\nu_{i}, \epsilon)+\mathcal{Q}(\nu_{i}, \epsilon)\right]
\end{eqnarray}

Similarly,  the methods can be applied to calculating $a_{2}$ in \eqref{A5}.
Doing so gives us the final result of $a_{2}$,
\begin{eqnarray}{\label{A13}}
-\frac{a_{2}}{4\pi^{2}}&=&\frac{a_{1}}{(4\pi^{2})^{2}}\Lambda^{2\epsilon}~\left[\left(\int \frac{d^{4}z^{-}_{1}d^{4}z^{+}_{2}d^{2}\theta_{2}d^{2}\bar{\theta}_{3}}{(X^{+}_{32})^{2}(X^{+}_{21})^{2(2-\epsilon)}}+permutations \right)
+\int\frac{d^{4}z^{-}_{1}d^{4}z^{+}_{2}d^{2}\bar{\theta}_{1}d^{2}\theta_{2}}{(X^{+}_{21})^{6-2\epsilon}}\right]\nonumber\\
&+&\frac{b_{1i}c_{i}}{4\pi^{2}}\Lambda^{2\epsilon-\nu_{i}}~\int
\frac{d^{4}z^{-}_{1}d^{4}x_{3}d^{4}\theta_{3}}{(X^{+}_{12})^{2-2\epsilon-\nu_{i}}(X^{+}_{13})^{2+\nu_{i}}(X^{+}_{32})^{2+\nu_{i}}}\nonumber\\
&+&4\times \frac{1}{(4\pi^{2})^{2}}\Lambda^{4\epsilon}~\int
\frac{d^{4}z^{-}_{1}d^{4}z^{-}_{3}d^{4}z^{+}_{4}d^{2}\bar{\theta}_{3}d^{2}\theta_{4}}{(X^{+}_{32})^{2(2-\epsilon)}(X^{+}_{14})^{2(2-\epsilon)}
(X^{+}_{34})^{2}}
\end{eqnarray}
The first integral in the first line of \eqref{A13}do not contributes,
while the second integral is the similar to \eqref{A9},
\begin{eqnarray}{\label{A20}}
\frac{a_{1}}{(4\pi^{2})^{2}}\Lambda^{2\epsilon}~\int\frac{d^{4}z^{-}_{1}d^{4}z^{+}_{2}d^{2}\bar{\theta}_{1}d^{2}\theta_{2}}{(X^{+}_{12})^{6-2\epsilon}}
=0
\end{eqnarray}
The second one can be simplifed by introducing the $\mathcal{I}(\nu_{i},\epsilon)$ function as in \cite{1203.5129},
which results in,
\begin{eqnarray}{\label{A14}}
b_{1i}c_{i}\Lambda^{2\epsilon-\nu_{i}}~\int
\frac{d^{4}z^{-}_{1}d^{4}x_{3}d^{4}\theta_{3}}{(X^{+}_{12})^{2-2\epsilon-\nu_{i}}(X^{+}_{13})^{2+\nu_{i}}(X^{+}_{32})^{2+\nu_{i}}}
\equiv -8\pi^{4}\frac{b_{1i}c_{i}}{\nu_{i}-2\epsilon}\mathcal{I}(\nu_{i},\epsilon)
\end{eqnarray}
with
\begin{eqnarray}{\label{I}}
\mathcal{I}(\nu_{i},\epsilon)=-\frac{(\nu_{i}-2\epsilon)(3-2\epsilon-\nu_{i})(2-2\epsilon-\nu_{i})}{8\pi^{4}}\Lambda^{2\epsilon-\nu_{i}}
\int \frac{d^{4}X^{+}_{23}d^{4}X^{+}_{31}}{(X^{+}_{23}+X^{+}_{31})^{4-2\epsilon-2\nu_{i}}(X^{+}_{23})^{2+\nu_{i}}(X^{+}_{31})^{2+\nu_{i}}}
\nonumber\\
\end{eqnarray}
The last integral in \eqref{A13}
\begin{eqnarray}{\label{A8}}
4&\times& \frac{1}{(4\pi^{2})^{2}}\Lambda^{4\epsilon}~\int
\frac{d^{4}z^{-}_{1}d^{4}z^{-}_{3}d^{4}z^{+}_{4}d^{2}\bar{\theta}_{3}d^{2}\theta_{4}}{(X^{+}_{32})^{2(2-\epsilon)}(X^{+}_{14})^{2(2-\epsilon)}
(X^{+}_{34})^{2}}\nonumber\\
&=&4\times \frac{1}{(4\pi^{2})^{2}}\Lambda^{4\epsilon}~\int
\frac{d^{4}X^{+}_{12}d^{4}X^{+}_{34}d^{4}X^{+}_{14}d^{2}\bar{\theta}_{13}d^{2}\theta_{42}}
{(X^{+}_{34}-X^{+}_{14}+X^{+}_{12}+2i\theta_{42}\sigma\bar{\theta}_{13})^{2(2-\epsilon)}(X^{+}_{14})^{4-2\epsilon}(X^{+}_{34})^{2}}
\equiv \frac{8\pi^{6}}{(4\pi^{2})^{2}\epsilon^{2}}\mathcal{T}(\epsilon)\nonumber\\
\end{eqnarray}
with
\begin{eqnarray}{\label{T}}
\mathcal{T}(\epsilon)=\frac{\epsilon^{2}(-2+2\epsilon)(-5+2\epsilon)}{2\pi^{6}}\Lambda^{4\epsilon}
\int \frac{d^{4}X^{+}_{12}d^{4}X^{+}_{34}d^{4}X^{+}_{14}}{(X^{+}_{34}-X^{+}_{14}+X^{+}_{12})^{6-2\epsilon}(X^{+}_{34})^{2}(X^{+}_{14})^{2(2-\epsilon)}}
\end{eqnarray}
after we change the integral variables $z^{-}_{1}\rightarrow X^{+}_{12}$,  $z^{-}_{3}\rightarrow X^{+}_{34}$,
 $z^{+}_{4}\rightarrow -X^{+}_{14}$,  $\bar{\theta}_{3}\rightarrow-\bar{\theta}_{13}$ ,  $\theta_{4}\rightarrow \theta_{42}$,
and use the equality $X^{+}_{32}=X^{+}_{34}-X^{+}_{14}+X^{+}_{12}+2i\theta_{42}\sigma\bar{\theta}_{13}$.

Consequently, we get the final expression of \eqref{A13}
\begin{eqnarray}{\label{15}}
a_{2}=16\pi^{4}\left(\frac{c^{2}_{i}}{\nu^{2}_{i}-4\epsilon^{2}}\right)\mathcal{I}(\nu_{i},\epsilon)-\frac{2\pi^{2}}{\epsilon^{2}}\mathcal{T}(\epsilon)
\end{eqnarray}\\
With the help of Mathematica, the functionals defined above can be evaluated.

\end{document}